# Electrical, Thermal and Spectroscopic Characterization of Bulk $Bi_2Se_3$ Topological Insulator

Rabia Sultana[1,2], Geet Awana[1,3], Banabir Pal[4], P.K. Maheshwari[1,2], Monu Mishra[1,2], Govind Gupta[1,2], Anurag Gupta[1,2], S. Thirupathaiah[4] and V.P.S. Awana[1,2]

[1]*National Physical Laboratory (CSIR), Dr. K. S. Krishnan Road, New Delhi-110012, India*
[2]*Academy of Scientific and Innovative Research, NPL, New Delhi-110012, India*
[3]*Department of Physics and Astrophysics Delhi University, New Delhi-110007, India*
[4]*Solid State and Structural Chemistry Unit, Indian Institute of Science, Bangalore, Karnataka, 560012, India*

**Abstract**
We report electrical (angular magneto-resistance, and Hall), thermal (heat capacity) and spectroscopic (Raman, x-ray photo electron, angle resolved photo electron) characterization of bulk $Bi_2Se_3$ topological insulator, which is being is grown by self flux method through solid state reaction from high temperature (950°C) melt and slow cooling (2°C/hour) of constituent elements. $Bi_2Se_3$ exhibited metallic behaviour down to 5K. Magneto transport measurements revealed linear up to 400% and 30% MR at 5K under 14 Tesla field in perpendicular and parallel field direction respectively. We noticed that the magneto-resistance (MR) of $Bi_2Se_3$ is very sensitive to the angle of applied field. MR is maximum when the field is normal to the sample surface, while it is minimum when the field is parallel. Hall coefficient ($R_H$) is seen nearly invariant with negative carrier sign down to 5K albeit having near periodic oscillations above 100K. Heat capacity ($C_p$) versus temperature plot is seen without any phase transitions down to 5K and is well fitted ($Cp = \gamma T + \beta T^3$) at low temperature with calculated Debye temperature ($\theta_D$) value of 105.5K. Clear Raman peaks are seen at 72, 131 and 177 cm$^{-1}$ corresponding to $A_{1g}^1$, $E_g^2$ and $A_{1g}^2$ respectively. Though, two distinct asymmetric characteristic peak shapes are seen for Bi $4f_{7/2}$ and Bi $4f_{5/2}$, the Se 3d region is found to be broad displaying the overlapping of spin - orbit components of the same. Angle-resolved photoemission spectroscopy (ARPES) data of $Bi_2Se_3$ revealed distinctly the bulk conduction bands (BCB), surface state (SS), Dirac point (DP) and bulk valence bands (BVB) and 3D bulk conduction signatures are clearly seen. Summarily, host of physical properties for as grown $Bi_2Se_3$ crystal are reported here.

## Introduction

Topological Insulators (TIs) is one of the most active, fascinating and challenging hot topic in condensed matter physics in recent years. In fact, the ongoing vast field of research leading to the discovery of some interesting TIs ($Bi_2Se_3$, $Bi_2Te_3$, and $Sb_2Te_3$) with unique properties and high application potential had attracted much attention of the condensed matter physicists worldwide [1-6]. TIs are electronic materials which behave like an ordinary insulator in the interior and conducting on the outside surface/edge. This dual nature i.e., insulating as well as conducting occurs due to the presence of bulk band gap in the interior and topologically protected conducting states on surface. The topological surface states in three dimensional (3D) TIs has been predominantly examined by angle resolved photoemission spectroscopy (ARPES), scanning tunnelling microscopy (STM) and band structure calculations [7-10]. The topologically protected conducting states in TIs arise due to the strong spin-orbit interaction (SOI) and time reversal symmetry (TRI). The TIs could be useful from application point of view as spintronics and quantum computation devices [1]. Interestingly, TIs also behave as superconductors when intercalated or doped with elements like Sr, Cu, Nb, Tl etc. [11-17].

Apparently, $Bi_2Se_3$ and $Bi_2Te_3$ are among the most studied bulk TI compounds [1-10], which crystallizes in a rhombohedral structure belonging to the $R\bar{3}m$ (D5) space group. It is well known that these bulk 3D TI exhibit gapless single Dirac cone at the surface coupled with strong SOI and a large bulk energy gap of 0.3eV between the bulk bands. [7-9]. Further, $Bi_2Se_3$ is a pure compound rather than an alloy like $Bi_xSb_{1-x}$ and the presence of relatively large bulk energy gap of 0.3eV holds promises for room temperature applications [10,19-20]. It is also known that $Bi_2Se_3$ exhibits a high thermoelectric figure of merit (ZT) for commercial applications [18]. The linear magneto-resistance (MR) as well as the angular dependent MR behaviour is observed in thin films, nano-plates, nano-wires/nano-ribbons, nano-sheets and crystals of different TIs like $Bi_2Te_3$ and $Bi_2Se_3$ [19,21–24]. Accordingly, $Bi_2Se_3$ is thought to be a promising TI system toward unique applications for next generation spintronics, quantum computation and optoelectronics.

In the present study, we investigate on the synthesis as well as characterization of bulk $Bi_2Se_3$ TI. Host of physical property characterizations including angular magneto-transport, heat capacity and hall measurement along with various spectroscopic studies viz. Raman, Photo electron and angle resolved photo electron are reported here for as grown $Bi_2Se_3$ topological insulator.

**Experimental details**
High quality bulk $Bi_2Se_3$ single crystals have been synthesized by self flux method through the conventional solid state reaction route [25, 26]. High purity (99.99%) Bismuth (Bi) and Selenium (Se) were used as the starting material. Stoichiometric amounts of starting materials were accurately weighed, well mixed and ground thoroughly inside the glove box (MBRAUN Lab star) under high purity Ar atmosphere to avoid oxidation of the sample. The homogeneously mixed powder was pressed into a rectangular pellet using the hydraulic press under a pressure of 50kg/cm$^2$ and then sealed into an evacuated (10$^{-3}$Torr) quartz tube. The vacuum encapsulated quartz tube was then sintered inside a tube furnace with step increase rate of 2˚C/minute up to 950˚C, kept there for 12 hours, then slowly cooled down to 650˚C at a rate of 2˚C/hour. Further, the furnace was switched off and allowed to cool to room temperature. Figure 1 shows the schematic heat treatment diagram. The resultant sample was taken out by breaking the quart tube carefully. The obtained sample was shiny and silver in colour which was mechanically cleaved for further measurements. Characterizations of the synthesized $Bi_2Se_3$ crystal were performed using X-Ray Diffraction (XRD), Raman spectroscopy, Scanning Electron Microscopy (SEM), and X-ray Photoelectron Spectroscopy (XPS). Further, the magnetic and resistivity measurements under magnetic field were carried out by a conventional four-probe method on a quantum design Physical Property Measurement System (PPMS) with fields up to 14Tesla.

**Results and discussion**
To examine the crystallographic structure, chemical composition and lattice parameters of the obtained sample, room temperature X-ray diffraction (XRD) was performed using Rigaku, MiniFlex-II X-ray diffractometer with Cu-Kα radiation (λ=1.5418 Å). Figure 2(a) shows the Rietveld fitted room temperature XRD pattern of the studied gently powdered bulk $Bi_2Se_3$ single crystal performed using the FullProf suite Toolbar. The compound is crystallized in rhombohedral structure within $R\bar{3}m$ (D5) space group. Apparently, the studied sample exhibits a single phase without any impurities within the XRD limits. The lattice parameters are a = b ≠ c and their corresponding values are: a = b = 4.14(2) Å and c = 28.701(3) Å respectively. The inset of figure 2(a) shows the single crystal XRD pattern from the surface of as obtained bulk $Bi_2Se_3$ crystal with 00l planes being aligned along one direction. It is clear from inset of figure 2(a) that the obtained crystals are textured and grown along c-direction.
Figure 2(b) shows the unit cell structure of the studied bulk $Bi_2Se_3$ single crystal as being formed and analysed by the VESTA software from the Rietveld refined lattice parameters and the atomic positions. The studied $Bi_2Se_3$ crystal exhibits a rhombohedral structure composed of three quintuple layers (QLs) stacked one over the other. As reported earlier, each QL arranged as Se-Bi-Se-Bi-Se is separated by weak van der Waals forces [1-6, 26]. Noticeably, this weak van der Waals force makes the QL to slide over each other similar to graphite and helps the single crystal to cleave easily.

The morphological characteristics as well as the chemical composition of the synthesized $Bi_2Se_3$ single crystal were analyzed by scanning electron microscopy (SEM) and Energy Dispersive X-Ray Spectroscopy (EDAX). The SEM image of the resultant bulk $Bi_2Se_3$ single crystal exhibited layered (00l) directional growth and the quantitative weight% of the atomic constituents (Bi and Se) was found to be near stoichiometric. Apparently, the studied crystal is pure (uncontaminated from impurities like Carbon or oxygen) composed of atomic constituents Bi and Se, details are reported by some of us in an earlier short article [25]. The temperature versus resistivity of the studied $Bi_2Se_3$ sample with and without various applied magnetic fields exhibited metallic behaviour down to 2K. Here, the field was applied in a direction perpendicular to the current flow. Accordingly, the ρ(T)H plots exhibited metallic behaviour having positive temperature coefficient values [25].

Figure 3(a – b) displays the percentage change of magneto resistance MR= [ρ (H)-ρ (0)]/ρ (0) as a function of applied magnetic field [H ⊥ c-axis (θ = 0˚) and H //c-axis (θ= 90˚)] upto 14Tesla at different temperatures (from 5K to 200K). Under perpendicular applied magnetic field (θ= 0˚) up to 14 Tesla, the studied $Bi_2Se_3$ crystal exhibits a linear positive MR value increasing from 60% at 200K to 400% at 5K [Figure 3(a)]. However, under parallel applied magnetic field (θ= 90˚) the MR value of the studied sample decreases from 30% at 5K to 6% at 200K [Figure 3(b)]. Clearly the MR is decreased by an order of magnitude in parallel applied magnetic field. Accordingly, we can say that the studied sample exhibited the largest, high field non - saturating positive MR (%) value reaching up to 400% at 5K, when measured under perpendicular magnetic field. The high value of MR (400%) seems interesting from both basic research as well as from application point of view. Thus, we observed a giant linear MR effect sensitive to the applied magnetic field (H) direction, which is maximum for perpendicular and least for parallel field.

Figure 3(c) represents the percentage change of magneto resistance MR= [ρ (H)-ρ (0)]/ρ (0) as a function of rotation angle (θ) under constant applied magnetic field and temperature of 10Tesla and 5K respectively. The peak is observed at θ=180˚ (perpendicular magnetic field), while the dips are seen at θ = 90˚ and θ =270˚ (parallel magnetic field).Thus, we can say that the resultant $Bi_2Se_3$ crystal has a unique response to the perpendicular component of magnetic field at different tilt angles, confirming the linear MR as a 2D magneto-transport effect as reported earlier [19].

Figure 4 displays the Hall coefficient ($R_H$) as a function of temperature ($T$) at constant magnetic field of 3Tesla. The $R_H$ is observed to increase with decreasing temperature down to 100 K along with exhibiting harmonic oscillations. Below 100K the curve tends to saturate. The inset of Figure 4 shows the Hall resistivity ($\rho_{xy}$) versus magnetic field (up to 5Tesla) plot at different temperatures from 330K to 5K of the studied $Bi_2Se_3$ crystal. Two important informations are drawn from Figure 4 and its inset that the $R_H$ is nearly invariant with temperature in range of 5K to 300K albeit with the near periodic oscillation above 100K and the carrier sign is negative althrough in the studied temperature range. In depth studies are underway to understand the Hall coefficient ($R_H$) behaviour, particularly above 100K.

Figure 5 displays the heat capacity ($C_p$) versus temperature ($C_p$ - $T$) plot for the studied $Bi_2Se_3$ crystal in temperature range of 250K down to 5K. The $C_p$ value increases with increase in temperature but, eventually saturates at higher temperature. The inset of figure 7 represents the $C_p/T$ versus $T^2$ plot of the studied crystal. Here, the $C_p$ is calculated using the formula $C_p = \gamma T + \beta T^3$, where $\gamma$ and $\beta$ represents the coefficients of Sommerfeld and phononic contribution respectively. The Debye temperature ($\theta_D$) value comes out to be 105.5K, calculated using the equation $\theta_D = [12\pi^4 N_A z k_B/5\beta]^{1/3}$ where, $N_A$ represents the Avogadro's constant, z is the number of atoms per formula unit and $k_B$ is the Boltzmann's constant. The nature of both the graphs is in agreement to the previously reported $Sb_2Te_3$ single crystal [27]. However, the intercept $\gamma$ crosses near to zero value. The reason may be that we measured $C_P$ down to 5K only, which had been measured and fitted down to 2K in ref. 27. None the less the shape of $C_p$ plot is identical to that as reported in ref. 27 and the values thus obtained are comparable.

To observe the vibrational modes of the studied $Bi_2Se_3$, Raman Spectroscopy of the synthesized crystal was recorded using the Renishaw Raman Spectrometer. Figure 6 shows the Raman spectra of bulk $Bi_2Se_3$ single crystal recorded at room temperature. According to group theory, the irreducible representation for zone centre phonon can be illustrated as $\Gamma = 2E_g + 2A1_g + 2E_u + 2A1_u$, where g and u represents the gerade (Raman active modes: $2E_g$, $2A1_g$) and ungerade (Infra red active modes: $2E_u$, $2A1_u$) modes respectively [28-30]. Figure 5 clearly shows three distinct Raman active modes at 72, 131 and 177 cm$^{-1}$ corresponding to $A_{1g}^1$, $E_g^2$ and $A_{1g}^2$ respectively. These values are in good agreement to the earlier reported results and affirm that the crystals are of good quality [29-30]. Consequently, we can say that these peaks are shifted to higher frequency in comparison to reported $Bi_2Te_3$ [26]. Interestingly, the other ($E_{1g}^1$) lowest frequency Raman active mode is not observed, which is similar to most of the earlier reported literature [31, 32].

Further, to examine the surface chemistry of the studied $Bi_2Se_3$ single crystal, X-ray photoelectron spectroscopy (XPS) was employed using Omicron multi-probe surface analysis system equipped with Monochromatic Al-K$\alpha$ source having excitation energy of 1486.7 eV. Figure 7(a-b) depicts the XPS spectra recorded for $Bi_2Se_3$ single crystal. Figure 7(a) represents the primary XPS region of Bismuth (Bi 4f). Bi 4f region exhibits two distinct asymmetric characteristic peak shapes for Bi $4f_{7/2}$ and Bi $4f_{5/2}$ with corresponding binding energies of 157.4eV and 162.8eV respectively. Accordingly, the difference in the binding energies appears to be 5.4eV, revealing the well separated spin - orbit components of Bi 4f region. Additionally, a blue shift of 0.4eV is also showed by the binding energies of Bi 4f peaks with respect to metallic bismuth [32,33]. Figure 7(b) displays the primary XPS region of Selenium (Se 3d). The observed Se 3d region is found to be broad and thus, de-convolution was performed to identify the peaks by Guassian Lorentzian function. Two different peaks were observed at Se $3d_{5/2}$ and Se $3d_{3/2}$ with binding energies of 53.0 eV and 53.8 eV respectively. Here, the difference in the binding energies is found to be 0.8 eV, displaying the overlapping spin - orbit components of Se3d peak. Further, the binding energies of Se3d peaks reveal a red shift of about 2.6eV with respect Se metal [32,33]. The elemental composition was calculated using the formula $\rho_{Se}/\rho_{Bi}$, where $\rho$ is the atomic density. Interestingly, the elemental composition value appears to be 1.27, which is close to the stoichiometric ratio of $Bi_2Se_3$.

Figure 8 depicts angle-resolved photoemission spectroscopy (ARPES) data of $Bi_2Se_3$. The measurements were performed at the APE beam line in Elettra Synchrotron, Trieste, equipped with a Scienta DA30 deflection analyzer. Fig.8 (a) schematically shows a typical measuring geometry in which *s*- and *p*-plane polarized lights are defined with respect to the analyzer entrance slit and the scattering plane (SP). Fig. 8 (b) depicts Fermi surface (FS) map measured using photon energy of 35 eV. Fig.8 (c) Shows energy distribution maps (EDMs) measured using photon energies of 35 eV and 45 eV from left to right, respectively along the cut shown by red dashed line in Fig. 8(b). All the data were measured using p-polarized light. In the left panel of (c), BCB, SS, DP and BVB represent bulk conduction bands, surface state, Dirac point and bulk valence bands, respectively. From the photon energy dependent EDMs as shown in Fig.8 (c), one can clearly notice that the bulk conduction and valance bands are sensitive to the used photon energy and thus, suggesting that these are of 3D character. On the other hand, surface states are intact with the varying photon energy. The ARPES results being obtained for the studied $Bi_2Se_3$ crystals are in general agreement with previous reports [4-11].

**Conclusion**

Summarily, in present article we reported on crystal growth, structure and in brief host of physical properties including electrical (angular MR), thermal (heat capacity and hall measurement) and spectroscopic (Raman, XPS and ARPES) for $Bi_2Se_3$ topological insulator.

**Acknowledgement**

Authors from CSIR-NPL would like to thank their Director NPL India for his keen interest in the present work. This work is financially supported by DAE-SRC outstanding investigator award scheme on search for new superconductors. Rabia Sultana thanks CSIR, India for research fellowship and AcSIR-NPL for Ph.D. registration. Authors thank Mrs. Shaveta Sharma for Raman studies. S.T. acknowledges support by DST, India through the INSPIRE-Faculty program (Grant No.: IFA14 PH-86). S.T. thanks Prof. D. D. Sarma for his enormous support in the department of SSCU, IISc. The authors thank P. Kumar Das and I. Vobornik for their support during the beam time at Elettra Synchrotron, Trieste.

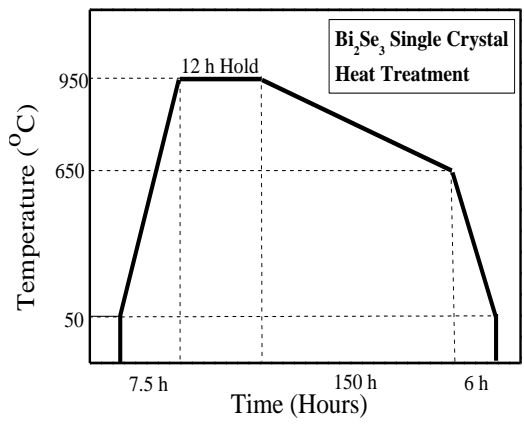
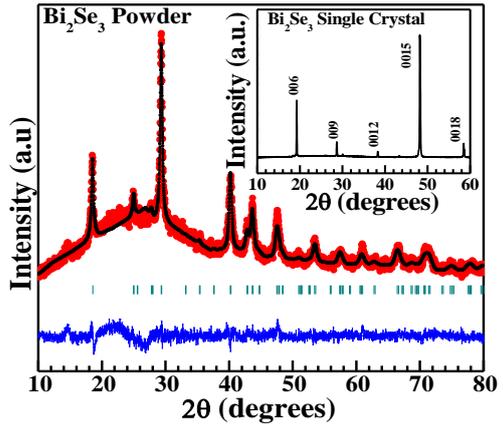

Figure 1                                   Figure 2(a)

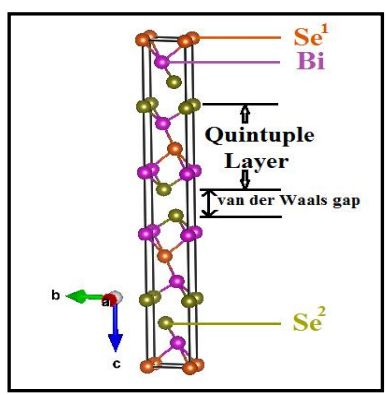
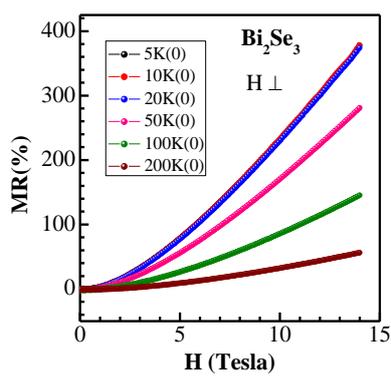
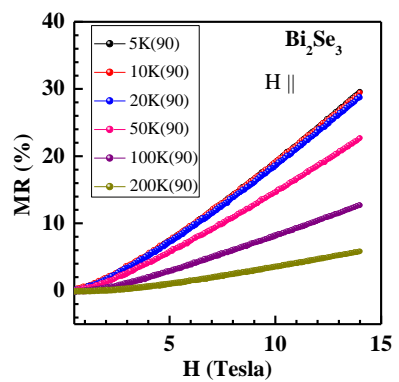

Figure 2(b)                    Figure 3(a)                    Figure 3(b)

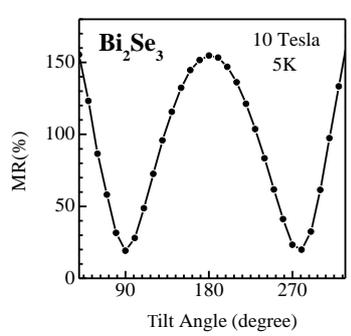
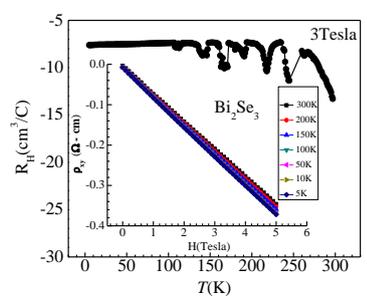
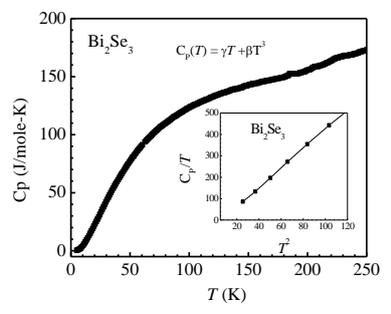

Figure 3(c)                    Figure (4)                    Figure (5)

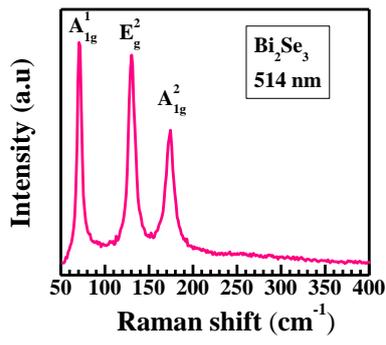 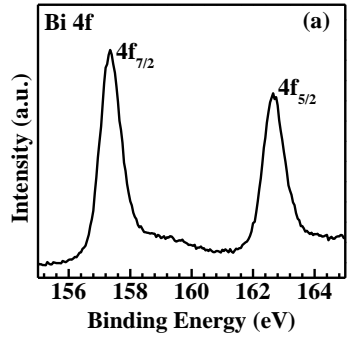 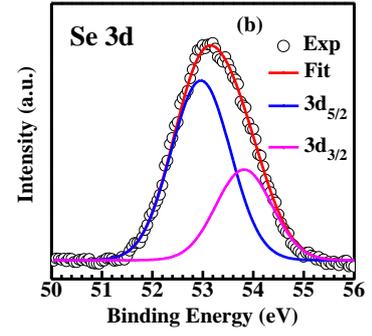

Figure (6)  Figure 7(a)  Figure 7(b)

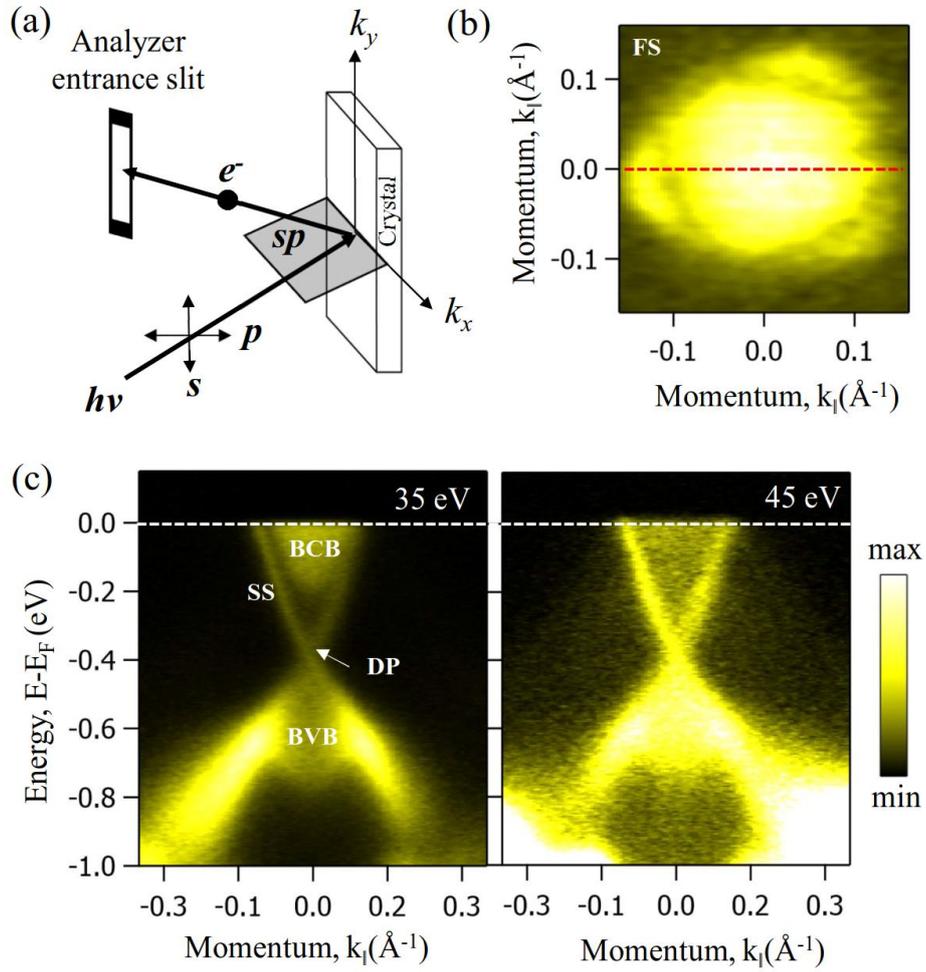

Figure 8